\documentstyle[prl,aps]{revtex}
\begin{document}
\draft
\twocolumn[\hsize\textwidth\columnwidth\hsize\csname @twocolumnfalse\endcsname

\title{Quasiparticle dispersion of the t-J and Hubbard models}

\author{A. Moreo, S. Haas, A. Sandvik, and E. Dagotto}

\address{Department of Physics and National High Magnetic Field Lab,
Florida State University, Tallahassee, FL 32306, USA}

\date{\today}
\maketitle

\begin{abstract}
The spectral weight ${\rm A({\bf p},\omega)}$ of the two
dimensional ${\rm t-J}$ and Hubbard models has been
calculated using exact diagonalization and quantum Monte
Carlo techniques, at several densities ${\rm 1.0 \leq \langle n \rangle \leq
0.5}$.
The photoemission $(\omega < 0)$
region  contains two dominant distinct features, namely
a low-energy quasiparticle peak with bandwidth of order J, and a
broad valence band peak at energies of order t. This behavior
$persists$ away from half-filling, as long as the antiferromagnetic (AF)
correlations are robust. The results give support to theories of the
copper oxide materials based on the behavior of holes in
antiferromagnets, and it also provides theoretical guidance for
the interpretation of experimental photoemission data for the cuprates.

\end{abstract}

\vskip2pc]
\narrowtext

Angle resolved photoemission (ARPES) techniques applied to the
high temperature superconductors have produced
interesting data that introduces important constraints on
theories for the copper oxide planes.
Recently, it has been shown\cite{flat-exper} that
the hole-doped compounds ${\rm Bi_2 Sr_2 Ca Cu_2 O_{8}}$,
${\rm Bi_2 Sr_2 Cu O_8}$, ${\rm Y Ba_2 Cu_3 O_7}$, and
${\rm Y Ba_2 Cu_4 O_8}$ exhibit
universal properties likely induced by the behavior
of carriers in their common ${\rm Cu O_2}$ planes. In particular, it has
been reported that the quasiparticle dispersion has a
small bandwidth governed by an
energy scale of the order of the exchange J of the Heisenberg
model (${\rm \sim 0.15 eV}$). In addition, in the vicinity of
momenta ${\rm Y = (0,\pi)}$ and ${\rm X = (\pi,0)}$, the
dispersion is anomalously flat.
These results give support to theoretical
ideas based on strongly correlated electrons, since (i) it is
well-established\cite{review} that
at half-filling the spectral function of a hole in an antiferromagnet
contains a sharp quasiparticle peak at the top of the valence band
spectra with
a bandwidth of order J, and
(ii) careful studies of the
fine details of the hole dispersion in one band models
have revealed the presence of
flat regions near the X and Y points in momentum
space.\cite{prl1,stratos,bulut1,hanke1}
The existence of
these two features is a direct consequence of the presence of
strong correlations and antiferromagnetism in the cuprates.

It is reasonable to assume that the behavior of holes in systems with
long-range antiferromagnetic order will not change qualitatively
as the density of holes is increased away from half-filling,
as long as the antiferromagnetic
correlation length $\xi_{AF}$ remains large.
Theories based on this
assumption have been proposed.\cite{spinbag,prl2} In particular, in
Refs.\cite{prl1,prl2} it was shown
that it is possible to reproduce many of the anomalous
properties of the cuprates, including the presence of a d-wave
superconducting state and the
existence of an optimal doping, with the economical
assumption that the sharp
quasiparticle peak observed at half-filling at the top of the valence
band remains robust as the electronic density decreases to
phenomenologically realistic
values. This assumption (i.e. approximate rigidity of the
quasiparticle dispersion with doping)
received support from recent
calculations addressing the presence of
``shadow bands'' in the cuprates.\cite{haas}
The rigid band hypothesis has also been studied by
other authors.\cite{trugman} On the experimental side,
recent ARPES results by Aebi et al.\cite{aebi} have shown that
features induced by the AF correlations at half-filling are also present
 at optimal doping. Since the closest structure to the Fermi level
in ${\rm A({\bf p},\omega)}$ is likely to dominate the
low temperature properties of the model,
then it is important to establish theoretically
whether the quasiparticle peaks
observed at half-filling
survive in  the presence of a finite density of holes.

The purpose of this paper is to report results of an extensive analysis
of the spectral weight for both
the 2D ${\rm t-J}$ and Hubbard models using exact diagonalization (ED)
and quantum Monte Carlo (QMC) methods, supplemented by Maximum
Entropy (ME)) techniques, and carried out at several densities.
${\rm A({\bf p}, \omega)}$ is shown to contain a two-peak
structure, with dispersing features near the top of the valence
band dominated by
the scale of antiferromagnetism J, while a secondary broad
structure appears at energies of order t.
We discuss the range in parameter space where this behavior
is to be expected, and its influence on the physics of carriers in
the cuprates.
However, note that
recent QMC results have
reported the presence of only $one$ PES
peak for the Hubbard model at both half-filling\cite{bulut2} and finite hole
density.\cite{bulut1}
We found that the disagreement with our present results is avoided once
the influence of finite temperature
effects is considered, and a more sophisticated ME method is used.

The technical details of the present computational study, as well as
the Hamiltonians of the Hubbard and ${\rm t-J}$ models, are
the standard ones,
unless otherwise stated.
In Fig.1a, ${\rm A({\bf p}, \omega)}$ is shown for the ${\rm t-J}$
model at $half-filling$ and ${\rm J/t=0.4}$
using the ED technique applied to
2D clusters with 16 and 18 sites. The combination of these
clusters allows enough resolution in momentum space to quantitatively
analyze the dispersion of the main features in the spectral weight.
The present
results have been obtained using approximately 100 iterations in the
standard continued fraction expansion (CFE) method
to obtain dynamical properties
using the Lanczos technique.\cite{review} However, the figure shows
that  only a small number of poles are
dominant. It is clear that near the Fermi energy, $(\omega=0)$,
there is a robust peak that weakly disperses in the scale of the figure.
Remnants of this low-energy peak exist at momenta $(0,0)$ and
$(\pi,\pi)$, in the latter barely visible to the eye (but its intensity
and position can be easily studied with the CFE approach mentioned above).
In Fig.1b, the position of the low-energy peak is shown with full dots,
with the convention that
the area of the dot is proportional to the intensity of the
peak. The bandwidth of this sharp quasiparticle-like peak is
${\rm \sim 0.8t=2J}$, in excellent agreement with our
expectations based on previous ED\cite{review} and Born
approximation\cite{stratos}
calculations. The flat region near $(\pi,0)$ is also
visible in the figure.
{}From Fig.1a it is clear that additional PES spectral weight in ${\rm A({\bf
p}, \omega)}$ is located at higher energies ${\rm |\omega|}$.
As discussed before in the literature, the strong correlation effects
force the hole quasiparticle to
carry only
a fraction of the integrated weight,\cite{bob1} and thus the presence
of considerable incoherent
intensity deep in energy
is reasonable. A rough
estimation of their position is shown in Fig.1b (open squares).\cite{string}
This feature is not relevant
for the low temperature behavior of the model which is
dominated by the quasiparticle peak at the top of the valence band.

Before describing the density dependence of our results,
let us clarify the importance of finite size effects in
Fig.1a,b, as well as the
differences between our results and those of
previous QMC simulations.\cite{bulut1,bulut2}
To address both issues simultaneously, we have carried out an extensive
QMC simulation of the 2D Hubbard model. The results reported here
correspond to ${\rm U/t=10}$ (i.e. the
strong coupling regime where the model should behave similarly to the
${\rm t-J}$ model), temperature
${\rm T=t/4}$, and using $\sim 10^5$ sweeps over the entire lattice to reduce
the
statistical errors. Here we use the
``classic'' ME technique.\cite{jarrell} This method gives
a closer fit to the Monte Carlo data than the variant used in
Ref.\cite{bulut1,bulut2} and therefore resolves more structure.
The  analytically-continued ${\rm A({\bf p}, \omega)}$
obtained at half-filling on an $8\times8$ cluster
is shown in Fig.2 at several momenta. The results
are both qualitatively and quantitatively
similar to those obtained for the ${\rm t-J}$ model in
Fig.1a,b, and also in good agreement with ED studies for the Hubbard
model.\cite{ortolani}
Two peaks in the PES region
are clearly identified for all momenta.
{}From their position it can be shown that
the bandwidth of the peak at the top of the valence band
is of order J, in excellent agreement with our previous discussion. The
second broader feature observed in the ED study of the ${\rm t-J}$ model is
also present in the QMC simulation results.
Studies at larger ${\rm U/t}$ couplings in the Hubbard model
and
in the region ${\rm 0.2 \leq J/t \leq 0.8}$ of the ${\rm t-J}$ model show
basically the same features. Then, here it is concluded
that the qualitative physics of both models
is very similar in the strong coupling region,
where a ${\rm A({\bf p}, \omega < 0)}$ with a double-peak
structure is observed,
as properly assumed in previous analytical studies.\cite{spinbag,prl1,comm2}

Let us now discuss our results away from half-filling. In the relevant regime
of density for the high-Tc superconductors, namely in the vicinity of
``optimal doping'' ${\rm \langle n \rangle \approx 0.85}$, the QMC+ME method at
large ${\rm U/t}$ produces stable results only at temperature
${\rm T=t/2}$ which is
too high to resolve the two peak structure even at half-filling. Thus,
in this density regime only the ED results are reliable.
In Fig.3a,b, ED data at density $\langle n \rangle
\approx 0.88$ are shown (two holes in the 16 and 18 sites clusters).
In this case $\xi_{AF}$ is approximately two lattice spacings.\cite{haas}
The PES results along the diagonal in momentum space
present structure very similar to that discussed at half-filling.
The low energy peak is well-defined at all momenta, even those
above the naive non-interacting Fermi momentum located
near $(\pi/2,\pi/2)$, and still it disperses with
a bandwidth of order J. The large accumulation of weight at
higher
energies ${\rm | \omega |}$ remains localized at $\omega \sim 4t$. Then, to the
extend
that the one band models reproduce the physics of the high-Tc cuprates,
it is reasonable to expect that PES experiments carried out at
half-filling $and$ near the optimal doping, should produce dispersive
features of similar intensity and bandwidth.
The clear similarity between the
experimental
bandwidth of the Bi2212 PES data, and recent results
for  the
$insulating$ ${\rm Sr_2 Cu O_2 Cl_2}$ compound,\cite{wells}
provides more evidence for the validity of
strongly correlated one band models for the cuprates. However, it
is important to remark that while the concrete prediction of our
calculations is that the bandwidth of the hole carriers is of order J,
the particular details of the dispersion may $differ$ from compound to
compound. For example, it has been recently remarked that
to reproduce the data for  ${\rm Sr_2 Cu O_2 Cl_2}$, the addition of
a small ${\rm t'}$-term to the 2D ${\rm t-J}$
model is necessary.\cite{gooding}
Thus, care must be taken when the fine details of
different compounds at different dopings are compared.

Now consider the inverse photoemission (IPES)
$(\omega > 0)$ intensity in Fig.3a,b.
The observed spectral weight in the vicinity of $(\pi,\pi)$
somewhat resembles the distribution for a
non-interacting Fermi system.
In principle, this effect does not seem
reproduced by a rigid band filling of
the states at half-filling. However, recently
Eder and Ohta\cite{eder} have shown that if proper
$quasiparticle$
operators\cite{bob1} are used in the calculation of the spectral
weight (i.e. operators dressed by spin fluctuations,
instead of bare electronic operators),
then the intensity of the IPES region is much reduced and the quality of
the rigid band description of the ${\rm t-J}$ model
appears more clearly. This is an important point not much emphasized in the
literature, namely that the robustness of the rigid band picture in
a given model $cannot$ be tested by analyzing the removal
of ``bare'' electrons (sudden approximation) as produced by
a PES experiment, but instead ``dressed'' carriers must be used. Thus,
PES and transport experiments may differ in their predictions if
holes are heavily renormalized as in the cuprates.

Fig.4a,b shows ED results for
${\rm A({\bf p}, \omega)}$ using the same clusters and
coupling as at half-filling, but now reducing further the density
to $\langle n \rangle \approx 0.75$ and $0.50$
(i.e. 4 and 8 holes in the 16 and 18 sites clusters). In this case,
through the spin correlations we
observed that $\xi_{AF}$ is less than
one lattice spacing and thus the influence of AF fluctuations
should be small at these densities. Indeed the two-peak structure
discussed before at higher densities
is now difficult to identify. While the broad
valence-band feature at $\omega \sim 4t$ remains,
only remnants of
the AF-induced intensity above the naive Fermi momentum can be
observed. The IPES signal
increased  its intensity and now ${\rm A({\bf p}, \omega)}$ resembles
the behavior of a
non-interacting ``$cosp_x + cosp_y$''
band.

An interesting detail of Figs.1a, 3a and 4a,b, is that
the intensity of PES weight at
${\bf p}=(\pi,\pi)$ changes appreciably as the density is varied.
This is to be expected since ${\bf p} = (\pi,\pi)$
is the momentum the most sensitive to the presence of AF correlations. In
particular, when $\xi_{AF} \rightarrow 0$, we expect that  the
PES weight at
${\rm {\bf p} = (\pi,\pi)}$
will be mostly transferred to the IPES regime. The presence of PES
weight at $(\pi,\pi)$ and $\langle n \rangle = 1$ is a direct
consequence of the AF correlations, and for a
paramagnetic background ${\rm A((\pi,\pi),\omega < 0)}$
should be negligible.

Summarizing, in this paper the quasiparticle dispersion of the 2D
${\rm t-J}$ and Hubbard models was analyzed as a function of the
electronic density. At half-filling, ${\rm A({\bf p}, \omega<0)}$
has a sharp quasiparticle-like peak at the top of the valence band with
a bandwidth of order J. This structure is the relevant one for the
low temperature behavior of the models.
A  second broad feature deeper in energy was also identified. As the
electronic density decreases, the ``two peak'' structure remains clearly
visible as long as the antiferromagnetic correlation length $\xi_{AF}$
is robust. When the AF fluctuations become negligible then a crossover
exists into a dispersion for the quasiparticles which resembles a
weakly interacting system.
For realistic values of the coupling, namely
${\rm U/t=10}$, this crossover from an antiferromagnetic metal to a
paramagnetic ground state occurs between $\langle n \rangle = 0.88$ and
$0.75$.
Then, in the interesting regime for the copper oxide materials
the AF correlations govern the behavior of the spectral weight.
The present results give strong
support to theories of the cuprates based on the
behavior of carriers in an antiferromagnet,\cite{spinbag,prl2} and
provides information about the crossover from a half-filled to a doped system
that
can guide the analysis of ARPES data.

After completing this work we received a preprint by Preuss, Hanke and
von der Linden\cite{hanke2} where conclusions similar to ours were reached.
E.D., A.M. and A.S. are supported by the Office of Naval Research under
grant ONR N00014-93-0495. E. D. is also supported
by the donors of the Petroleum Research Fund
administered by the American Chemical Society.
S. H. is supported by SCRI, at FSU.
We thank the NHMFL and MARTECH
at FSU for its partial support.


\vfil\eject

%
%
{\bf Figure Captions}
\begin{enumerate}

\item
(a) Spectral weight ${\rm A({\bf p}, \omega)}$ of the 2D ${\rm t-J}$ model
at ${\rm J/t=0.4}$
using clusters of 16 and 18 sites along the
diagonal in momentum space. The $\delta$-functions have been given
a width $\epsilon = 0.25t$ in the plots; (b) position of the
two dominant peaks in ${\rm A({\bf p}, \omega)}$ as a function
of momentum. The area of the circles is proportional to the intensity
of the quasiparticle peak they represent. The error bars denote the
width of the peak as observed in Fig.1a (sometimes to a given broad peak
several poles contribute appreciably).
The full squares at $\omega \sim -4t$ represent the center of the broad
valence band weight, and the area of the squares is $not$ proportional
to their intensity.

\item
Spectral weight ${\rm A({\bf p}, \omega)}$ of the 2D Hubbard model
obtained with the QMC method supplemented by Maximum-Entropy, on
an $8 \times 8$ cluster, ${\rm U/t=10}$, and ${\rm T=t/4}$.

\item
Same as Fig.1 but for density $\langle n \rangle \approx 0.88$
(i.e. two holes on the 16 and 18 sites clusters). In (a) the
PES intensity is shown with a solid line, while the IPES
intensity is given by a dotted line. The chemical potential is
located at $\omega = 0$. In (b) the full and open circles
represent the PES and IPES intensities, respectively, of the peaks
the closest to the Fermi energy. Their area is proportional to
the intensity.

\item
(a) Same as Fig.3a but for density $\langle n \rangle \approx 0.75$
i.e. 4 holes on the 16 and 18 sites clusters; (b) Same as Fig.3a but
for density $\langle n \rangle \approx 0.50$ i.e. 8 holes on the 16 and
18 sites clusters.

\end{enumerate}


\begin{references}
\bibitem{string} The third structure immediately after the quasiparticle peak
has been extensively analyzed under the name of ``string-states''.\cite
{review,stratos}
These features should be more sensitive to doping than the dominant
quasiparticle state, and thus they are not discussed in detail here.
\bibitem{comm2}
The simulations of Ref.\cite{bulut1,bulut2} did not
observe such structure mainly due to finite temperature effects and
also the use of a different ME technique with less resolution.
\bibitem{comm1}
This result is to be contrasted with ${\bf p} = (0,0)$.
The PES weight at this momenta is not
as drastically density-dependent as at $(\pi,\pi)$ because
for both AF and paramagnetic backgrounds the $(0,0)$ state is
populated at all finite densities.
\end{references}
\end{document}